\documentclass[draft]{agujournal2019}
\usepackage{url} 
\usepackage{lineno}
\usepackage[finalnew]{trackchanges} 
\usepackage{soul}
\draftfalse

\journalname{Geophysical Research Letters}

\begin{document}

%
%

\title{A comprehensive picture for binary interactions of subaqueous barchans}

%
%

\textcolor{blue}{An edited version of this paper was published by AGU. Copyright 2020 American Geophysical Union.\\
Assis, W. R. and Franklin, E. M. (2020). A comprehensive picture for binary interactions of subaqueous barchans. Geophysical Research Letters, 47, e2020GL089464, DOI 10.1029/2020GL089464.\\
To view the published open abstract, go to https://doi.org/10.1029/2020GL089464.}




\authors{W. R. Assis\affil{1}, E. M. Franklin\affil{2}}


\affiliation{1,2}{School of Mechanical Engineering, UNICAMP - University of Campinas,\\
Rua Mendeleyev, 200, Campinas, SP, Brazil}




\correspondingauthor{Erick M. Franklin}{franklin@fem.unicamp.br}




\begin{keypoints}
\item We identify five binary interactions of barchans and propose classification maps
\item We show \change{experimental evidence}{indications} that an ejected barchan has the same mass of the impacting one
\item We found that the asymmetry of the downstream dune is large in wake-dominated processes
\end{keypoints}

%
%

%
%


\begin{abstract}
We investigate experimentally the short-range interactions occurring between two subaqueous barchans. The experiments were conducted in a water channel of transparent material where controlled grains were poured inside, and a camera placed on the top acquired images of the bedforms. We varied the grain types (diameter, density and roundness), pile masses, transverse distances, water flow rates and initial conditions. As a result, five different patterns were identified for both aligned and off-centered configurations and we propose interaction maps that depend basically on the ratio between the number of grains of each dune, Shields number and alignment of barchans. In addition, we show experimental \change{evidence}{indications} that an ejected barchan has roughly the same mass of the impacting one in some cases, and that in wake-dominated processes the asymmetry of the downstream dune is large. The present results shed light on the size regulation of barchans found on Earth and other planets.
\end{abstract}

\section*{Plain Language Summary}
Barchans are crescent-shaped dunes that are often organized in dune fields, where binary interactions and collisions play a significant role in regulating their dynamics and sizes. Barchan collisions  are frequent in many environments, such as Earth's deserts and on the surface of Mars, but their large time scales (the decade and the millennium for aeolian and Martian collisions, respectively) compared to the aquatic case (of the order of the minute) make subaqueous barchans the ideal object of study. Taking advantage of that, we performed experiments in a water channel of transparent material, where pairs of barchans were transported by the water flow while a camera acquired images of them. We found five different types of barchan-barchan interaction, and propose maps that provide a comprehensive classification for the short-range interactions of subaqueous barchans. In addition, we show that, in some cases, an ejected barchan has roughly the same mass of the impacting one, and that the perturbation of the flow caused by the upstream barchan generates large asymmetries in the downstream one. Our results represent a significant step toward understanding the barcanoid forms and size regulation of barchans found in water, air, and other planetary environments.

\section{Introduction}
The interaction between a fluid flow and a granular bed gives rise to different kinds of bedforms. Of particular interest are the crescent-shaped dunes, called barchans, that are formed under one-directional flow and limited amount of available grains, being encountered in different environments such as rivers, water ducts, Earth's deserts and other planetary environments \cite{Bagnold_1, Herrmann_Sauermann, Hersen_3, Elbelrhiti, Claudin_Andreotti, Parteli2}. Although barchans may grow as isolated bedforms \cite{Alvarez, Alvarez3}, they are often organized in dune fields, where dune-dune interactions play a significant role in regulating their dynamics and sizes \cite{Hersen_2, Hersen_5, Kocurek, Genois, Genois2}.

Over the past decades, several studies investigated the collisions and short-distance interactions of aeolian barchans based on field measurements \cite{Norris, Gay, Vermeesch, Hugenholtz}. Yet, because these measurements are based on aerial images, the time series for barchan collisions are usually incomplete given the long timescales of aeolian interactions (of the order of the decade), hindering a comprehensive understanding of barchan collisions. Because of their much faster scales (of the order of the minute), some studies investigated the interactions of barchans in water flumes and tanks \cite{Endo2, Hersen_5}, from which different collision patterns were identified and their dynamics described. In addition, numerical simulations using continuum \cite{Schwammle2, Duran2, Zhou2} and simplified discrete models \cite{Katsuki} could reproduce some of the collision patterns, shedding light on the essential mechanisms involved. However, the simplifications present in those models precluded them from reproducing correctly all barchan interactions, failing to predict the correct split of dunes in some cases and predicting soliton behavior in others.

By observing that a solitary barchan within a dune field is marginally stable, tending to grow or shrink once the stable size is disturbed, and the existence in nature of corridors of barchans, \citeA{Hersen_2} proposed a model for the formation of corridors, and \citeA{Hersen_5} showed that barchan collisions could be important for the size regulation of barchans. In order to better understand the mechanisms behind the formation of corridors with size-selected barchans, \citeA{Duran3} and \citeA{Genois2} introduced simplified models based on sand flux balances and elementary rules for barchan collisions. With such models, \citeA{Duran3} showed that collisions are important for size regulation and inter-barchan spacing, while \citeA{Genois2}, by adjusting sand fluxes, obtained corridors of sparse and large or dense and small barchans according to the balance between sand fluxes and collision types, showing that sand distribution due to collisions organizes barchans in narrow corridors of size-selected dunes. \citeA{Bo}, based on numerical simulations using a scale-coupled model, found that the probability of barchan collisions varies with the flow strength, grain diameter, grain supply and height ratio of barchans. They quantified the probabilities for the occurrence of three different types of barchan collisions within a dune field (merging, exchange and fragmentation-exchange, described next), but not how the collision processes vary with the considered parameters.

Although many previous studies were devoted to barchan collisions, the problem is still not completely understood and a general picture is lacking. The emerging patterns, though present in both aeolian and aquatic environments, have not yet had all their important parameters identified, so that universal expressions or maps for predicting the results of collisions do not exist. The identification of collision patterns from the approaching of subaqueous barchans until the end of the collisional process was performed by \citeA{Endo2} in the case of aligned dunes for different mass ratios, and by \citeA{Hersen_5} in the case of off-centered dunes for different transverse distances of centroids of colliding dunes (referred to as impact or offset parameter), while \citeA{Bo} focused on the probabilities of barchan collisions in a dune field obtained from numerical computations. However, how the diameter, density and roundness of grains, flow strength and initial conditions affect the collision patterns remains to be investigated. In addition, mass transfers between barchans during collisions are not completely understood.

In this Letter we investigate extensively the binary interactions, including binary collisions, of subaqueous barchans. We carried out exhaustive measurements of the short-range interactions between two barchan dunes\add{, i.e., when their longitudinal separation is of the order of the size of the upstream bedform,} by varying the mass of initial piles, their alignment (centered or off-centered)\add{, initial longitudinal separation}, \remove{the }grain properties (diameter, density and roundness), flow strength and initial conditions (downstream barchan already formed or to be developed), most of them affecting the patterns emerging from interactions. We identify five types of binary interactions for both aligned and off-centered barchans, and show \change{evidence}{indications} that an ejected barchan has roughly the same mass of the impacting one in cases involving collisions with exchange of grains and that in wake-dominated processes the asymmetry of the downstream dune is large. We propose a new classification for the binary short-range interactions of subaqueous barchans that depends on the ratio between the number of grains of each dune, Shields number and barchans alignment, shedding light on the size regulation of barchans in a dune field.

\section{Materials and Methods}

The experimental device consisted of a water reservoir, two centrifugal pumps, a flow straightener, a 5-$m$-long closed-conduit channel of transparent material and rectangular cross section (width = 160 mm and height 2$\delta$ = 50 mm), a settling tank, and a return line. A pressure-driven flow was imposed in the channel by means of the centrifugal pumps, and the flow followed the order just described. The channel test section was 1 m long and started 40 hydraulic diameters, 40 $\times$ $d_h$, downstream of the channel inlet, assuring a developed channel flow just upstream the bedforms, where $d_h$ = 3.05 $\delta$ is the cross-sectional area multiplied by four and divided by the cross-sectional perimeter. With the channel previously filled with water, controlled grains were poured inside, forming a pair of bedforms in either aligned or off-centered configurations. By imposing a water flow, each bedform was deformed into a barchan shape and interacted with each other. We used different initial conditions, in which we placed a first pile and let it deform into a barchan dune before placing an upstream pile, or we let it deform in half-way a barchan dune before placing the second pile, or we placed two conical piles and let them deform together into barchan dunes, and the mass ratio of the piles, defined here as the mass of the upstream pile (impacting) divided by that of the downstream one (target), varied within 0.005 and 1. The initial longitudinal distance between bedforms was of the order of the diameter of the upstream pile\add{, $D$, being within 0.22 and 3.6$D$ between dune borders (smaller distance between dunes in the longitudinal direction),} and, because the dune velocity varies with the inverse of its size \cite{Bagnold_1}, the mass of the impacting dune was always equal or lesser than that of the target dune. A camera placed above the channel acquired images of the bedforms and, therefore, we did not measure systematically the barchan height. However, based on reported values of the aspect ratio of barchans \cite{Andreotti_1} and our experimental observations, we estimate the crest heights as approximately 5 mm, i.e., 10 \% of the channel height. The layout of the experimental device, a photograph of the test section, and microscopy images of the used grains are shown in the supporting information.

A total number of \change{113}{123} tests were performed, for which we used tap water at temperatures within 22 and 30 $^{\circ}$C and different populations of grains (not mixed): round glass beads ($\rho_s$ = 2500 kg/m$^3$) with $0.15$ mm $\leq\,d\,\leq$ $0.25$ mm and $0.40$ mm $\leq\,d\,\leq$ $0.60$ mm, angular glass beads with $0.21$ mm $\leq\,d\,\leq$ $0.30$ mm, and zirconium beads ($\rho_s$ = 4100 kg/m$^3$) with $0.40$ mm $\leq\,d\,\leq$ $0.60$, where $\rho_s$ and $d$ are, respectively, the density and diameter of grains. The cross-sectional mean velocities of water, $U$, varied between 0.226 and 0.365 m/s, corresponding to Reynolds numbers based on the channel height, Re = $\rho U 2\delta /\mu$, within $1.13$ $\times$ $10^4$ and $1.82$ $\times$ $10^4 $,  where $\mu$ is the dynamic viscosity and $\rho$ the density of the fluid. The shear velocities on the channel walls (base state), $u_*$, were computed based on measurements with a two-dimensional particle image velocimetry (2D-PIV) device \cite{Franklin_9, Cunez2, Alvarez3} and found to follow the Blasius correlation \cite{Schlichting_1}, being within 0.0133 and 0.0202 m/s. By considering the fluid velocities applied to each grain type, the Shields number, $\theta = (\rho u_*^2)/((\rho_s - \rho )gd)$, varied within 0.019 and 0.106, where $g$ is the acceleration of gravity (see supporting information for a description of the PIV tests, estimated deviations in $u_*$ and $\theta$, and lists of all tested conditions).

\section{Results}

\begin{figure}
    \begin{center}
     \includegraphics[width=.99\linewidth]{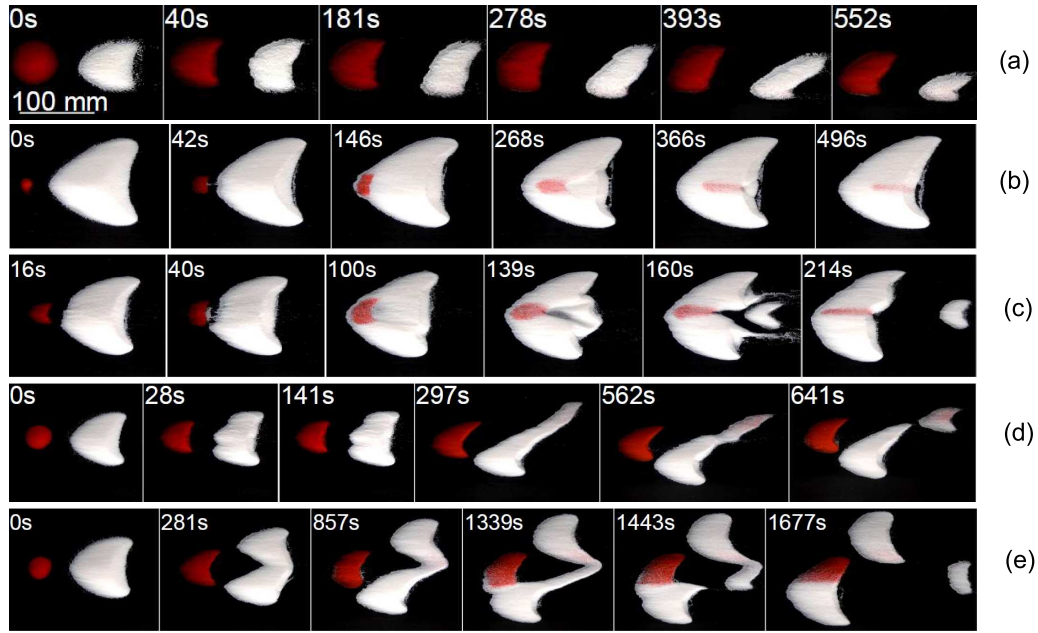}
    \end{center}
\caption{Snapshots of barchan interactions for aligned dunes. In the snapshots, the water flow is from left to right, the upstream pile consisting of red (darker) glass beads and the downstream pile of white (clearer) glass beads, and the corresponding times are shown in each frame. In Figure (a), $0.40$ mm $\leq\,d\,\leq$ $0.60$ mm and in the remaining figures $0.15$ mm $\leq\,d\,\leq$ $0.25$ mm. (a) Chasing; (b) merging; (c) exchange; (d) fragmentation-chasing; (e) fragmentation-exchange, and they correspond to test numbers 61, 65, 36, 5 and 22 in the table of Fig. S23 of the supporting information, respectively.}
	\label{fig:snapshots_align}
\end{figure}

\begin{figure}
    \begin{center}
     \includegraphics[width=.99\linewidth]{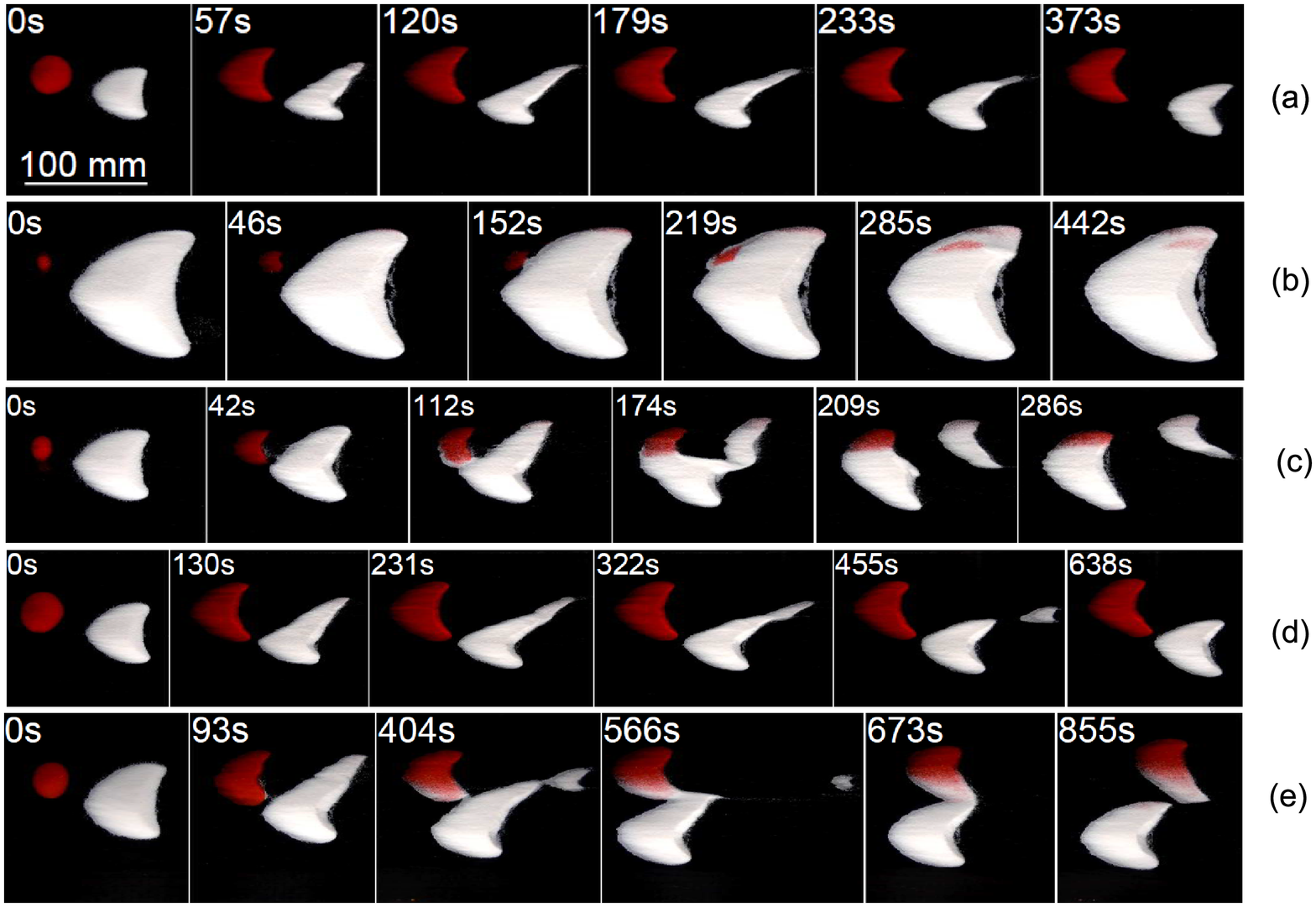}
    \end{center}
\caption{Snapshots of barchan interactions for off-centered dunes. In the snapshots, the water flow is from left to right, the upstream pile consisting of red (darker) glass beads and the downstream pile of white (clearer) glass beads of $0.15$ mm $\leq\,d\,\leq$ $0.25$ mm, and the corresponding times are shown in each frame. (a) Chasing; (b) merging; (c) exchange; (d) fragmentation-chasing; (e) fragmentation-exchange, and they correspond to test numbers 43, 38, 41, 31 and 5 in the table of Fig. S24 of the supporting information, respectively.}
	\label{fig:snapshots_stag}
\end{figure}

Five different patterns were observed as resulting from the short-range interaction, as can be seen in Figures \ref{fig:snapshots_align} and \ref{fig:snapshots_stag}, that show, respectively, snapshots of barchan interactions for the aligned and off-centered cases: 1) chasing (Figures \ref{fig:snapshots_align}a and \ref{fig:snapshots_stag}a), when the upstream dune does not reach the downstream one. This pattern appears when the barchans have almost the same size, and the wake of the upstream dune, by increasing turbulent levels and creating channeling \cite{Palmer, Bristow}, promotes a larger erosion on the downstream dune, which then shrinks and moves faster (even if it receives grains from the upstream barchan); 2) merging (Figures \ref{fig:snapshots_align}b and \ref{fig:snapshots_stag}b), when the upstream dune reaches the downstream one and they merge; 3) exchange (Figures \ref{fig:snapshots_align}c and \ref{fig:snapshots_stag}c), when, once the upstream dune reaches the downstream one, a small barchan is ejected and, being the smaller one, outruns the other and migrates downstream. The first impression is that the impacting barchan traverses the target one, but the use of marked grains shows that there is mass exchange between them, as can be seen in Figures \ref{fig:snapshots_align}c and \ref{fig:snapshots_stag}c. In some cases, depending on the sum of sizes of the impacting and target barchans, the ejected barchan is so small that it is close to the minimum size \cite{Franklin_8} and spreads out just after being ejected; 4) fragmentation-chasing (Figures \ref{fig:snapshots_align}d and \ref{fig:snapshots_stag}d), when, due to the wake of the upstream dune \cite{Palmer, Bristow}, in particular just downstream the reattachment point of the recirculation region, the downstream dune splits before being reached. Because the divided dunes are smaller than the upstream one, they outrun the upstream dune; and 5) fragmentation-exchange (Figures \ref{fig:snapshots_align}e and \ref{fig:snapshots_stag}e), when fragmentation as in (4) initiates, but, the upstream barchan being still faster then the splitting dune, the former reaches the latter. Once they touch, an off-center exchange occurs, and a small barchan is ejected. In the aligned configuration, the ejected barchan results from the interaction of an elongated horn with the remaining of the divided dunes, while in the off-centered configuration the ejected barchan is the smaller portion of the splitting dune. Finally, this redistribution of grains having finished, the smaller dunes are downstream and, therefore, three resulting barchans migrate without reaching each other. Movies showing all the five dune-dune interactions for both configurations and snapshots for other grain types are available as supporting information.

The presence of the five patterns in both aligned and off-centered configurations shows that variations of the offset (or impact) parameter, although influencing the conditions where patterns can appear, are not crucial for their appearance. Also, the mass ratio alone cannot regulate the appearance of all collision patterns, \citeA{Endo2} and \citeA{Duran2} having not found the five patterns for aligned dunes by varying only their mass ratio. \citeA{Endo2} identified only the merging, exchange and fragmentation-chasing patterns (which they named absorption, ejection and split), and \citeA{Duran2}, based on numerical simulations, the merging and exchange patterns (which they called coalescence and breeding), but the latter with a different behavior than our experimental observations. In addition, they found a pattern called budding, which could be equivalent to the fragmentation-exchange, but, in fact, is different, the target dune splitting only after the collision had happened, and also a solitary wave behavior, which is not observed experimentally. However, until now the different patterns emerging from collisions have been described in terms of only the offset parameter and mass ratio \cite{Katsuki2, Genois2, Genois}.

We observed in our experiments that, in addition to these parameters, the fluid shearing and mass of each grain are also of importance, the latter, combined with the pile masses, being equivalent to the number of grains forming the piles. If, in one hand, the difference in the number of grains (or, also, the mass ratio) gives the time scale for collision, on the other hand the total number of grains (or the sum of pile volumes) gives the total size of the system, indicating if the resulting barchan is too large, with tendency to split because of instabilities of hydrodynamic nature \cite{Andreotti_1, Andreotti_2, Charru_3, Franklin_12}. In addition, the flow strength and the size and density of grains are also related to hydrodynamic instabilities and to minimum sizes regulating the wavelength of bedforms and favoring the split of dunes or even their spread out \cite{Andreotti_2, Parteli, Franklin_8, Charru_5, Courrech}, so that they also must be taken into account. For example, \citeA{Khosronejad} showed that new barchans can be generated by a calving process on the horns of existing barchans, caused by the fluid shearing on the horn surface. Therefore, barchan collisions would be better described by the number of grains forming each pile, size and density of grains, flow strength and alignment of barchans, instead of only the mass ratio of piles and the offset parameter. Another aspect not investigated in previous studies is the effect of initial conditions of bedforms on barchan collisions (target barchan being initially a fully-developed barchan, a partially-developed barchan, or a conical pile). For the initial conditions, as well as the grain roundness, we did not observe any significant difference in our experiments (see supporting information for snapshots of barchan interactions with two conical piles as initial condition).

We propose that the short-range interaction patterns can be described by the offset parameter, the Shields number, and the number of grains forming each pile. For the latter, the difference in the number of grains forming each pile, $\Delta_N$, is proportional to the relative velocity of dunes, while the sum of those numbers, $\Sigma_N$, is proportional to the total size of the bedform once dunes have collided. We then introduce the dimensionless particle number:

\begin{equation}
\xi_N = \frac{\Delta_N}{\Sigma_N}
\end{equation}

\begin{figure}
	  \begin{center}
     \includegraphics[width=.95\linewidth]{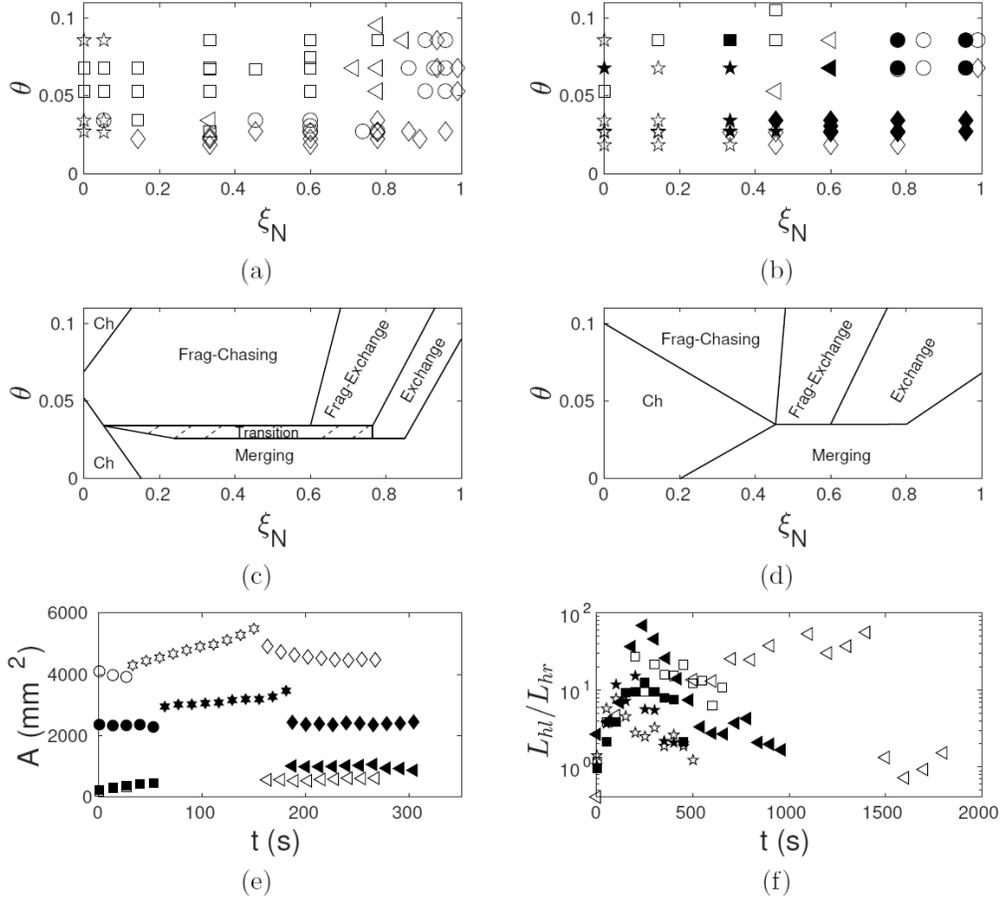}
    \end{center}
\caption{Figures (a) and (b): Patterns of barchan-barchan interactions as functions of $\xi_N$ and $\theta$ for (a) aligned and (b) off-centered barchans. Stars, diamonds, circles, squares and triangles correspond to chasing, merging, exchange, fragmentation-chasing and fragmentation-exchange, respectively. In Figure (b), open symbols correspond to $\sigma$ $<$ 0.5 and solid symbols to $\sigma$ $\geq$ 0.5. Figures (c) and (d): Boundaries between different patterns for the aligned and off-centered barchans, respectively, where Ch stands for chasing and Frag to fragmentation. Figure (e): Area variation along time for the exchange pattern. Squares and circles correspond to the initial upstream (impact) and downstream (target) barchans, respectively, stars to the merged bedform, and diamonds and triangles to the merged bedform and new (expelled) barchan, respectively (tests 36 of Fig. S23 and 41 of Fig. S24 of the supporting information). Note that open squares are difficult to visualize in the graphic because they are at the same positions of solid squares. Fig (f): Ratio between the lengths of the left and right horns, $L_{hl}/L_{hr}$, of the downstream dune along time. Stars, squares and triangles correspond to chasing, fragmentation-chasing and fragmentation-exchange patterns, respectively (tests 61, 5 and 22 of Fig. S23, and 43, 31 and 5 of Fig. S24 of the supporting information). In Figs (e) and (f), open symbols correspond to the aligned and solid symbols to off-centered cases. All individual images that were processed to plot Figures (e) and (f) are available on Mendeley Data (http://dx.doi.org/10.17632/jn3kt83hzh\remove{.1})}
	\label{fig:maps}
\end{figure}

\begin{sloppypar}
\noindent The Shields number is the ratio between mobile and resisting forces, linked to the fluid shearing and the grain weight, respectively, so that it takes into account the flow strength and the size and density of grains. Finally, the alignment of barchans is represented by the offset parameter $\sigma$ (dimensionless), computed here as the transverse distance between the centroid of approaching barchans, $\eta$, divided by their average width: $\sigma$ = $2\eta / (W_U + W_D)$, where $W_U$ and $W_D$ are the widths of the upstream and downstream bedforms, respectively, and $\eta$ is positive to the left of the target dune (with respect to the flow direction). Although we recognize three dimensionless groups, we decided to present all our data in two 2D maps in order to organize the patterns in the simplest and comprehensive way that we could find. Therefore, we plotted one interaction map for the aligned case (Figure \ref{fig:maps}a) and another one for the off-centered case (Figure \ref{fig:maps}b), where patterns are shown as functions of $\xi_N$ and $\theta$. In addition, for the off-centered case the map is parametrized by $\sigma$ $<$ 0.5 or $\sigma$ $\geq$ 0.5, which indicates if the the offset is relatively small or large, respectively. The number of grains forming each pile was considered as the ratio between the pile and grain masses.
\end{sloppypar}

Figures \ref{fig:maps}a and \ref{fig:maps}b show that the interaction patterns are relatively well organized by the $\xi_N$ and $\theta$ groups\add{, independent of the initial longitudinal separation between bedforms}, with transition regions between them where patterns are sometimes difficult to classify (their behavior in these regions is close to two patterns). Conscious of this difficulty, we drew lines separating the different patterns, which we present in Figures \ref{fig:maps}c and \ref{fig:maps}d for the aligned and off-centered configurations, respectively. We drew such lines based solely on the experimental observations, and they consist in a tentative way to classify the different patterns in $\theta$ vs. $\xi_N$ maps. Although computation of those lines based on stability analyses or other analytical method remains to be done, we believe that the present maps may be useful for predicting the output of short-range barchan-barchan interactions under different conditions.

Based on image processing, we tracked the bedforms along the acquired images for each of the five patterns and identified some of their characteristic lengths. Because of approximately constant ratios between barchan dimensions \cite{Hersen_1, Andreotti_1}, the projected area of a developed barchan multiplied by its height (around 10\% its width) is proportional to its volume, and, therefore, to its mass. However, in the present case barchans are being formed and deformed, interacting with each other, so that those relations are not completely valid. Conscious of that, we decided to analyze the projected areas of barchans as an indicator of the quantity of grains forming the dunes. Figure \ref{fig:maps}e presents the instantaneous values of the projected area of bedforms along time for the exchange pattern, and Figure \ref{fig:maps}f the evolution of the ratio between the lengths of the left and right horns (with respect to the flow direction), $L_{hl}$ and $L_{hr}$, respectively.

We start by observing that the area of the upstream bedform increased in the beginning of all experiments because it was initially a conical pile, with a higher ratio between its height and length, and, therefore, it spread out once the water flow was imposed. While the upstream barchan was growing, the downstream one was already formed and lost grains by its horns without receiving much grains from the upstream bedform, so that its area decreased slightly in the beginning of each test. Figure \ref{fig:maps}e shows also that the area of the dune resulting from the collision increases along time due to its spreading, since just after collision the upstream dune (impact dune) climbs over the downstream one (target dune), as can be seen on movies available as supporting information. After that, a new born barchan is expelled with roughly the same area of the impact dune \add{(see supporting information for a table showing the areas of impacting and expelled dunes of 15 tests)}. This indicates that the mass of the generated barchan is approximately that of the impacting one, though the constituent grains are not the same (Figures \ref{fig:snapshots_align}c and \ref{fig:snapshots_stag}c). Although this mass exchange of same value has been conjectured before \cite{Vermeesch}, being even confounded with a solitary behavior in some cases \cite{Schwammle2}, it had never been \change{measured in}{assessed from} controlled experiments until now.

Finally, from Figure \ref{fig:maps}f we observe experimental evidence that the asymmetry of the downstream dune is large in wake-dominated processes (i.e., when the growth of one of the horns is due mainly to the fluid flow), the asymmetry being lower in the case of collision-generated asymmetries (not shown in Figure \ref{fig:maps}f, but presented in the supporting information). This implies that the wake of upstream dunes \cite{Palmer, Bristow}, and not the collision itself, generates most of horns asymmetries. Although the origin of horns asymmetries has been studied previously \cite{Parteli4}, it needs to be investigated further in the specific case of dune-dune interactions.

Although our experiments were limited to the subaqueous case, the resulting analysis may be useful for predicting barchan-barchan interactions in other environments, such as Earth's deserts and on the surface of Mars. However, we expect differences related to the larger quantities of grains involved in the aeolian and Martian dunes and, in particular, the trajectories followed by individual grains according to the state of the fluid. For the trajectories, grains move mainly by rolling and sliding and follow closely the fluid flow in the subaqueous case, being susceptible to small vortices and other small structures of the flow. This has been shown to be especially important for the grains migrating to the barchan horns \cite{Alvarez3, Alvarez4}. When the fluid is a gas, grains move by saltation and reptation, and those in saltation follow basically a straight line in the main flow direction \cite{Bagnold_1}, being undisturbed by the small structures of the flow. Discrepancies between the present analysis and the behavior in other environments are likely to occur, mainly for the wake-dominated processes, but where and when they occur, and to what extent, remain to be investigated.

\section{Conclusions}
In conclusion, subaqueous barchan-barchan interactions result in five different patterns for both aligned and off-centered configurations, being well organized in two maps as functions of $\xi_N$ and $\theta$ and parametrized by $\sigma$. These maps provide a comprehensive and simple classification for the short-range interactions of subaqueous barchans and, although we have not analyzed the binary collisions in Earth's deserts and other planetary environments, given their long timescales, they may be useful for predicting the collisions of barchans in different environments. The present results represent a significant step toward understanding the barcanoid forms, barchan asymmetries and size regulation of barchans found in water, air, and other planetary environments.

\acknowledgments
\begin{sloppypar}
W. R. Assis is grateful to FAPESP (grant no. 2019/10239-7), and E. M. Franklin is grateful to FAPESP (grant no. 2018/14981-7), to CNPq (grant no. 400284/2016-2) and to FAEPEX/UNICAMP (grant no. 2112/19) for the financial support provided. Data supporting this work are available in the supporting information and in Mendeley Data (http://dx.doi.org/10.17632/jn3kt83hzh\remove{.1}).
\end{sloppypar}

\bibliography{references}

\end{document}